\documentclass[aip,jmp,amsmath,amssymb,reprint]{revtex4-1}
\usepackage[utf8]{inputenc}

\usepackage{amsfonts}
\usepackage{eucal}
\usepackage{bm}
\usepackage{tensor}
\usepackage{xfrac}
\usepackage{mathtools}
	\DeclarePairedDelimiter\parentheses{\lparen}{\rparen}
	
\usepackage{ragged2e}
\usepackage{longtable}
\usepackage{color}
\usepackage{hyperref}

\newcommand{\rhovec}{\bm{\rho}}
\newcommand{\qvec}{\bm{q}}
\newcommand{\Omegavec}{\bm{\Omega}}

\newcommand{\Gammaoz}{\Gamma_{\kern-1pt\mathrm{oz}}}
\newcommand{\Gammat}{\bm{{\Gamma}}}
\newcommand{\jvec}{\bm{j}}
\newcommand{\dstar}{\mathrm{d}^{\kern-1pt*}}

\newcommand{\U}{\mathcal{U}}

\newcommand{\F}{\bm{{\mathcal{F}}}}
\newcommand{\G}{\bm{{\mathcal{G}}}}

\newcommand{\entmix}{\xi}
\newcommand{\rbar}{\overline{r}}
\newcommand{\tbar}{\overline{t}}

\newcommand{\Gammar}{\overline{\Gamma}}

\newcommand{\vcd}{\dot{v}_c}
\newcommand{\vdd}{\dot{v}_d}
\newcommand{\vfd}{\dot{v}_f}

\DeclareMathOperator{\arctanh}{atanh}

\begin{document}

\author{Rico A.\,R.\ Picone}
\email[Email: ]{rpicone@stmartin.edu}
\affiliation{Saint Martin's University, Department of Mechanical Engineering, Lacey, USA}
\affiliation{University of Washington, Department of Mechanical Engineering, Seattle, USA}
\affiliation{Institute for Soldier Healing, Seattle, USA}
\author{John A.\ Sidles}
\affiliation{University of Washington, Department of Mechanical Engineering, Seattle, USA}
\affiliation{Institute for Soldier Healing, Seattle, USA}
\author{Joseph L.\ Garbini}
\affiliation{University of Washington, Department of Mechanical Engineering, Seattle, USA}
\title{Application of a thermodynamical framework for transport analysis\\ to the derivation of Dirac's value function}
\affiliation{Institute for Soldier Healing, Seattle, USA}
\begin{abstract}
		From a non-equilibrium thermodynamical framework for transport analysis based on Onsager's Regression Hypothesis, we derive the value function first described by Dirac for isotope separation. This application of the framework is interpreted as both further validation of the transport framework and as a generalization of Dirac's value function.
		The framework for the analysis of transport phenomena is introduced, first.  
		From the entropy of mixing, and in the presence of gradients in thermodynamic potentials, this framework generates a dynamical transport model from which Dirac's value function is derived as a measure of separative work performed. 
		Dirac's value function is thus shown to be a measure of separative work for systems that are described by the entropy of mixing.
		As a further demonstration of its generality, the result is applied to a two-quantity, single spatial-dimension spin magnetization system.
\end{abstract}

\maketitle


\section{Introduction}

\noindent This work arose with specific applications in mind: magnetic resonance imaging technologies such as clinical MRI, nuclear magnetic resonance (NMR) spectroscopy, and magnetic resonance force microscopy (MRFM).
These technologies share a common challenge: greater imaging speed and resolution require greater sample polarization (i.e.\ magnetization). Hyperpolarization techniques based on dynamic nuclear polarization (DNP) have made strides in recent years, and show great promise.\cite{Isaac2016,Ni2013,Krummenacker2012,Abragam1978} 
The authors set out to develop an alternative hyperpolarization method called separative magnetization transport (SMT).\cite{Picone2015,Picone2014b} Unlike DNP, the SMT mechanism is based on spin transport and therefore required the development of a non-equilibrium thermodynamical model of transport. 
In order for the model to be sufficiently general and flexible (not to mention consistent with the laws of thermodynamics), it proved necessarily to develop a mathematically rigorous framework for transport analysis. 
In principle, this framework can be applied to many types of separative systems, including those of isotope separation, thermoelectric separation, and magnetization separation. Classic results of magnetization transport (Genack and Redfield's dynamical equations and the Langevin paramagnetic equation), and one of isotope separation (the Fenske equation), were derived from the framework,\cite{Picone2015} but---until now---there had been no direct connection to one of the fundamental equations of isotope separation: Dirac's value function (DVF).

Dirac's value function is a key measure of the separative 
work required to concentrate some desired substance, such as a valued isotope,
from a dilute feedstock.\cite{Dirac:1995aa} Despite the economic and strategic importance of Dirac's value function, Dirac's own derivation was never published in the
open literature.  Although textbooks present plausible arguments for its
efficacy,\cite{Cohen1951,Bernstein2009} to our knowledge no published derivation describes a class of systems, broader than those of isotope separation, whose entropy function is dominated by the entropy of mixing.


We begin the derivation of Dirac's value function by motivating it from physical considerations, illustrating each general consideration by concrete considerations of isotope separation.  Suppose that we have a cascade of devices (diffusion chambers or centrifuges) that each accept a single input stream (the \emph{feed} stream) and generate two output streams (the \emph{product} stream and the \emph{waste} stream).  Suppose further that the separative capacity of the device is governed by a transport model (which is derived in this article) that assumes the separating device is subject to the entropy of mixing.  Can we assign to the output streams of the device a value function, such that the separative work performed by the device is independent of the concentration of the input stream?  The answer is ``yes'' and the unique function so defined is the DVF.

Section~\ref{sec:framework} introduces the framework, and---from this---Section~\ref{sec:model} derives a transport model for systems described by two conserved quantities, one spatial dimension, and the entropy of mixing. From the transport model and the definition of separative work, Section~\ref{sec:DVF} derives the DVF. Given this new derivation of the DVF, Section~\ref{sec:generality} summarizes its general aspects. Finally, an application to a spin magnetization system is presented in Section~\ref{sec:application}.


\section{Introduction to a Framework and a Model of Transport}\label{sec:frameworkandmodel}

Previous derivations of Dirac's value function have been limited in two respects: 
	(1) they have been confined to isotope separation systems and 
	(2) they have not explicitly considered the transport currents driving the separation. 
Consequently, we begin by presenting a framework for transport analysis in Section~\ref{sec:framework}, from which a dynamical transport model based on the entropy of mixing is derived in Section~\ref{sec:model}. 
This model describes a transport current from which Dirac's value function can be derived, as shown in Section~\ref{sec:DVF}. 
Therefore, both limitations of previous derivations are mitigated: the first by the generality retained in the dynamical transport model; the second by deriving Dirac's value function from the model's transport current.
Furthermore, this study provides additional verification of the generality of the framework presented in this section.

\subsection{General Framework of Transport} \label{sec:framework}

A differential geometric framework for transport based on Onsager's Regression Hypothesis has already been presented by the authors.\cite{Picone2014b,Picone2015} In overview, a dynamic transport model for a given system is generated by specifying the system's conserved quantities, entropy density function, physical geometry, and transport rates. The framework is described below, then applied to separative transport in Section~\ref{sec:model}.

A Riemannian manifold $\U$ represents the physical space within which conserved quantities $\qvec \in \mathbb{R}^n$ have spatial densities $\rhovec \in \mathcal{O}^*$, where $\mathcal{O}^*$ is the set of smooth maps from $\U \times \mathbb{R}$ (where $\mathbb{R}$ represents time) to $V^* = \mathbb{R}^n$ (that is, for each point in space and time we assign a vector in $\mathbb{R}^n$); i.e.,
\begin{align}
	\qvec = \int_\U \rhovec\ dv
\end{align}
where $dv$ is a volume element of $\U$. The spatial manifold $\U$ has Riemannian metric $g = g_{\alpha\beta}\ dr^\alpha \otimes dr^\beta$, where $r$ is the local spatial coordinate. The standard thermodynamic dual\footnote{It is perhaps physically natural to consider the quantities $\rhovec$ to be dual to the potentials $\Omegavec$, as we have done here, although the choice is a matter of convention and the converse is also valid.} basis $(\varepsilon^1,\ldots,\varepsilon^{n})$ for $V^*$ is
\begin{align*}
	\varepsilon^1 = (1,0,\ldots,0),\
	\ldots,\
	\varepsilon^{n} = (0,0,\ldots,1).
\end{align*}
In terms of this basis, the conserved quantity densities are written as
\begin{align}
	\rhovec = [\rho_\varepsilon]_i\ \varepsilon^i,
\end{align}
where the Einstein summation convention is adopted (as it is throughout the following).

Let $s: V^* \rightarrow \mathbb{R}$ be a local entropy density function that is nonnegative and concave. We use the Legendre transform relationship\cite{Zia2009} to define the local thermodynamic potentials $\Omegavec:\U \times \mathbb{R} \rightarrow V$ in terms of the exterior derivative\footnote{The exterior derivative operator $d$ is here taken with respect to the vector space $V^*$.} $d$ to be $\Omegavec = ds \circ \rhovec$. The standard thermodynamic basis for $\Omegavec$ is $(E_i)$, where $E_i(\varepsilon^j) = \delta_i^j$, so we write in components
\begin{align}\label{eq:Omega}
	\Omegavec = [\Omega_E]^i E_i.
\end{align}
Thermodynamic potentials $\Omegavec$ represent the chemical potentials of the substance.\cite{Landau2013}


With the local spatial coordinates $(r^\alpha)$ on $\U$, the tangent space at a point $p \in \U$ has coordinate vectors $\partial/\partial r^\alpha|_p$; the tangent bundle is $T\U$. Similarly, the cotangent space at point $p$ has coordinate covectors $dr^\alpha|_p$; the cotangent bundle is $T^*\U$.

We now define the current $\jvec$ to be a gradient of a thermodynamic potential. Equations of this form incldue Fick's first law of diffusion,\cite{Giddings1991} Ohm's law of electrical potential and current,\cite{Feynman2011,Giddings1991} Darcy's law of hydraulic flow,\cite{Whitaker1986} and Fourier's law of thermal conductivity. The current is
\begin{align}\label{eq:j}
	\jvec &= \F \circ (d \Omegavec)^\sharp,
\end{align}
where $\F$ is a positive-definite thermometric tensor field that acts as a thermodynamic and spatial metric (and will be described in detail below), and $\sharp$ is the ``sharp'' musical isomorphism (see Appendix~\ref{section: musical} for references to musical notation). The exterior derivative $d$ is here taken with respect to spatial coordinates.

Conservation is expressed by a continuity equation
\begin{align} \label{eq:continuity}
	\partial_t \rhovec = \dstar\jvec
\end{align}
where $\dstar$ is the Hodge codifferential operator (see \emph{e.g.} Lee\cite{Lee:2012ab}). \autoref{eq:continuity} is the governing dynamical equation of the framework.

So far, the three physical insights required to define the transport dynamics are:
\begin{enumerate}
	\item the identification of conserved quantities $\qvec$, 
	\item the identification of the spatial manifold $\U$, and
	\item the selection of an entropy density function $s$.
\end{enumerate}
%

There remains undefined only the thermometric tensor field $\F$, which incorporates both the spatial metric $g$ and a thermodynamic metric we will denote $\G$. Moreover, $\F$ incorporates transport rates via the mixed tensor field $\Gammat$. The definition of $\F$, then, is
\begin{align}\label{eq:F}
	\F = -\Gammat \circ g \circ \G^{-1}.
\end{align}
We can identify $\F$ to be a generalized (tensor field) version of the kinetic coefficients described by Lars Onsager.\cite{Casimir1945,Onsager1953,Onsager1931,Yablonsky2011}

The symmetry of $\F$ is asserted by the Onsager-Casimir Reciprocal Relations,\cite[Sections 2.4\,\&\,11.2]{Lebon2010} which are grounded in Onsager's Regression Hypothesis.\cite[Section 8.2]{Chandler1987}

Let us consider $\Gammat$, the transport rate tensor field. It is system-dependent in that it describes the transport rates among the conserved quantity densities $\rhovec$. This is the fourth and final physical insight required. If a given system has a single transport rate for all quantity densities, then $\Gammat$ contains only a single nonzero component, so it is effectively a scalar. 

Finally, let us consider $\G$, the thermodynamic metric. We use a form of the Ruppeiner metric,\cite{Ruppeiner1979, Ruppeiner1995} which allows us to specify a thermodynamic metric from the entropy density function:
\begin{align}\label{eq:G}
	\G =
		\frac{\partial^2 s}{\partial \rho_i \partial \rho_j}\ E_i \otimes E_j.
\end{align}
This metric is the foundation of the framework. Without it, the dynamics of the system could not be derived from the entropy density function $s$.

This concludes the introduction to the framework for transport.  Note that it can be applied to a system having any number of conserved quantities.   

\subsection{Transport Model}\label{sec:model}

Many thermodynamic systems can be described by just two conserved quantities, of which energy is one. Examples of such systems include an thermoelectric systems (energy and charge are conserved), isotope separation systems (energy and mass are conserved), and nuclear spin magnetization (energy and magnetic moment are conserved). In the context of spin magnetization transport, the framework has been used to derive a two-quantity transport model.\cite{Picone2014b,Picone2015} Here we generalize those results with a view toward deriving a value function.

Conserved quantities are coupled to a long-range potential field, such as a gravitational, electrical, or magnetic potential, and commonly these potentials are spatially varying.  In the following development, the conserved quantities are coupled to a spatially varying field $\phi(r)$. 

We adopt a standard basis in which the two components of the quantity densities $\rhovec$ describe two independently conserved densities, of which the first component $[\rho_\varepsilon]_1$ is the total energy density and the second component $[\rho_\varepsilon]_2$ is the transported quantity density that couples to $\phi(r)$. 

Examples of such pairs include
\begin{enumerate}
	\item the total electrical energy density $[\rho_\varepsilon]_1$ of charges in an electrical field and charge density $[\rho_\varepsilon]_2$,
	\item the total energy density $[\rho_\varepsilon]_1$ (kinetic and potential) of massive particles in a gravitational field and the volumetric spatial density $[\rho_\varepsilon]_2$ of the particles, and
	\item the total magnetic energy density $[\rho_\varepsilon]_1$ (dipole and Zeeman) of spins in a magnetic field and spin magnetization $[\rho_\varepsilon]_2$.
\end{enumerate}

For simplicity, we assume that transport occurs in a single Euclidean spatial dimension, such that the Euclidean metric is $g = dr \otimes dr$. 

\subsubsection{Thermodynamic Bases}\label{sec:bases}

To introduce an entropy function, thermodynamic bases must be considered.  In addition to the standard thermodynamic bases $(\varepsilon^i)$ and $(E_i)$, a spatially inhomogeneous basis for which the dynamical equations greatly simplify is now introduced. The following transformation defines this thermodynamic basis:
\begin{align}
	[\rho_e]_i= [P]\indices{_i^j} [\rho_\varepsilon]_j,
\end{align}
where $P$ is defined as a mixed tensor with matrix representation
\begin{align}\label{eq:P1}
	P(r) = 
	\begin{bmatrix}
		1/\rho_{1,\text{m}} &
		-\phi(r)/\rho_{1,\text{m}} \\
		0 &
		1/\rho_{2,\text{m}}
	\end{bmatrix},
\end{align}
where $\rho_{1,\text{m}}$ and $\rho_{2,\text{m}}$ are the maximum possible values of the quantity densities, which normalize both $[\rho_e]_1$ and $[\rho_e]_2$ such that they map to the interval $[-1,1]$.\footnote{Some quantities can be negative, whereas others will use only the interval $[0,1]$.} This basis transformation effectively subtracts the potential energy due to the external field $\phi(r)$ from the total energy, leaving only the local dynamical energy density $[\rho_e]_1$. For instance, a massive particle in a gravitational field would have total energy density (kinetic + potential) $[\rho_\varepsilon]_1$ and local dynamic energy density (total - potential = kinetic) $[\rho_e]_1$. The transformation also has a normalizing effect on $[\rho_\varepsilon]_2$ such that $[\rho_e]_2$ is a normalized transport quantity density. The new covector basis is called the inhomogeneous basis and is denoted $(e^i)$. The inhomogeneity is in its spatially varying basis covectors. The covector basis $(e^i)$ has a dual basis $(e_i)$, defined by the requirement that $e^i(e_j) = \delta_i^j$, where $\delta_i^j$ is the Kronecker delta function.

The transformation $P$ has the following motivations: (1) it removes the explicit influence of potential energy, (2) it simplifies the expressions that follow, and (3) it presents the components of $\rhovec$ and $\Omegavec$ in a dimensionless, normalized form. Although useful, it is not a fundamentally necessary transformation because $\bm{\rho} = [\rho_e]_i e^i = [\rho_\varepsilon]_i \varepsilon^i$. However, we will see that, although the inhomogeneity of the \emph{basis} is not essential, the inhomogeneity of the external \emph{field} $\phi(r)$ is essential for separation to occur (see Section~\ref{sec:physicalinsights}).

\subsubsection{Entropy Density Function} \label{sec:entropy}

For a normalized variable $x \in (-1,1)$, the entropy of mixing $\entmix: (-1,1) \rightarrow \mathbb{R}$ can be expressed as\cite{Denbigh1981}
\begin{align}
	\entmix(x) = \frac12 \ln{4} 
	&+ \frac12 \left(x - 1\right) \ln{\left(1 - x\right)} 
	- \frac12 \left(x + 1\right) \ln{\left(1 + x\right)}.
\end{align}
Entropy is an extensive property, and therefore the entropy of each quantity can be summed.\cite{Prigogine1968} Therefore, let the local entropy density function $s: V^* \rightarrow \mathbb{R}$ be defined as
\begin{align}\label{eq:s}
	s(\rhovec) = \entmix([\rho_e]_1) + \entmix([\rho_e]_2),
\end{align}
This formulation includes contributions to the entropy of mixing from both the local dynamical energy density and the normalized transport quantity density. The zero of entropy is inconsequential for the model because the framework is invariant to the zero of entropy.\cite[pp.~15-16]{Picone2014b}

\subsubsection{Transport Rate Tensor Field}

For two conserved quantities and one spatial dimension, the general transport rate tensor field $\Gammat$ has four entries. Separate transport rates are assumed for each quantity, such that
\begin{align}
	\Gammat &=
	[\Gamma_e]_1^1 \left( e^1 \otimes dr \right) \otimes \left( e_1 \otimes dr \right) + \nonumber \\
	&\phantom{{}={}} [\Gamma_e]_2^2 \left( e^2 \otimes dr \right) \otimes \left( e_2 \otimes dr \right).
\end{align}
Therefore, we assign only two transport rates.

\subsubsection{Dimensionless Spatial Coordinate, Temporal Variable, and Transport Rate}\label{sec:dimensionless}

It is convenient to use dimensionless variables. The conserved quantity densities can be expressed in a dimensionless manner by changing to the inhomogeneous basis of Section~\ref{sec:bases}. Space, time, and rate constants are now transformed such that they are dimensionless for the analysis that follows. We define the dimensionless spatial coordinate by $\rbar = a\,r$ and the dimensionless time variable by $\tbar = [\Gamma_e]_2^2 a^2 t$, where $a$ is a system-dependent length constant.

The transport rates are made dimensionless by dividing each by $[\Gamma_e]_2^2$. This can be seen by considering the simple transport continuity equation of a single variable in diffusion: $\partial_t \rho = \Gamma \partial_r^2 \rho$. Transforming this into the dimensionless spatial coordinate and time variable, we obtain
\begin{align}
	\partial_{\tbar} = \partial_{\rbar}^2 \rho.
\end{align}
A similar procedure gives the dimensionless transport rates for the two-quantity case to be
\begin{alignat}{3}
	\Gammat &= \phantom{{}+{}}
	\Gammar &&\left( e^1 \otimes d\rbar \right) \otimes &&\left( e_1 \otimes d\rbar \right) \nonumber \\
	&\phantom{{}={}} + 1 &&\left( e^2 \otimes d\rbar \right) \otimes &&\left( e_2 \otimes d\rbar \right),
\end{alignat}
where $\Gammar \equiv [\Gamma_e]_1^1/[\Gamma_e]_2^2$.

\subsubsection{Thermodynamic Potentials}

Thermodynamic potentials $\Omegavec: \U \times \mathbb{R} \rightarrow V$ are related to the quantity densities $\rhovec: \U \times \mathbb{R} \rightarrow V$ by the Legendre transform, as described in Section~\ref{sec:framework}. For the entropy density function of \autoref{eq:s}, we find
\begin{align}\label{eq:Omegarho}
	\Omegavec = -\arctanh{[\rho_e]_1}\ e_1 - \arctanh{[\rho_e]_2}\ e_2.
\end{align}
Conversely, it can be shown that
\begin{align}
	\rhovec = -\tanh{[\Omega_e]_1}\ e^1 - \tanh{[\Omega_e]_2}\ e^2.
\end{align}
These relationships among the thermodynamic potentials and quantity densities allow us to easily transform variables.

\subsubsection{Ruppeiner Metric and the Kinetic Coefficient Tensor Field}

The form of the Ruppeiner metric defined in \autoref{eq:G} evaluates to
\begin{alignat}{2}
	\G = 
	&\left( ([\rho_e]_1)^2 - 1 \right)^{-1} \ e_1 &&\otimes e_1 \ + \nonumber \\
	&\left( ([\rho_e]_2)^2 - 1 \right)^{-1} \ e_2 &&\otimes e_2.
\end{alignat}
The inverse of this metric approaches zero as the quantities tend toward their maximum of unity. This will have an important effect on the transport, as will become apparent. We can now write the kinetic coefficient tensor field, defined in \autoref{eq:F},
\begin{alignat}{3}
	\F = \
	&\Gammar &&\left( 1 - ([\rho_e]_1)^2\right) &&\ e^1 \otimes e^1 \otimes d\rbar \ + \nonumber \\
	\ &1 &&\left( 1 - ([\rho_e]_2)^2 \right) &&\ e^2 \otimes e^2 \otimes d\rbar.
\end{alignat}
This tensor field acts as both a thermodynamic and a spatial metric, as well as sets the transport rates.

\subsubsection{Transport Current}\label{sec:j}

The transport current $\jvec$ is an important quantity and will play a crucial role in the derivation of the value function of Dirac in Section~\ref{sec:DVF}. \autoref{eq:j} yields the expression
\begin{alignat}{3}\label{eq:j2}
	\jvec = &- \Gammar\, \partial_{\rbar} [\rho_e]_1 \ e^1 \otimes d\rbar \ + \nonumber \\
					&\left( c \left( 1 - ([\rho_e]_2)^2\right) \arctanh{([\rho_e]_1)} - \partial_{\rbar} [\rho_e]_2 \right)  \ e^2 \otimes d\rbar
\end{alignat}
where, for notational convenience, we are defining the function $c$ to be the dimensionless quantity
\begin{align}
	c(\rbar) = \frac{\rho_{2,\text{m}}}{\rho_{1,\text{m}}} \cdot \frac{d\phi}{d\rbar}.
\end{align}
We will now consider some implications of this form of the transport current.

\subsubsection{Physical Insights from the Transport Current}\label{sec:physicalinsights}
Let us consider the consequences of the external field $\phi(r)$ being uniform:
\begin{align}
	d\phi/d\rbar = 0 \quad 
	\Rightarrow  \quad c(\rbar) = 0  \quad
	\Rightarrow  \quad \nonumber \\
	\jvec = - \Gammar\, \partial_{\rbar} [\rho_e]_1 e^1 \otimes d\rbar - \partial_{\rbar} [\rho_e]_2  e^2 \otimes d\rbar.
\end{align}
The current has only diffusive terms, so no separative transport can occur. In fact, the term
\begin{align}
	c \left( 1 - ([\rho_e]_2)^2\right) \arctanh{([\rho_e]_1)}\nonumber
\end{align}
is the only term that contributes to separative transport. The second factor $\left( 1 - ([\rho_e]_2)^2\right)$ is a direct contribution of Ruppeiner's metric. If the polarization approaches unity, separative flow is ``throttled'' by decreasing entropy. The final factor shows that the energy density $[\rho_e]_1$ also affects the separative current.

\subsubsection{Continuity Equation}

Although we will use \autoref{eq:j2} in the derivation of the value function of Dirac, the model would be incomplete without the inclusion of the continuity equation. It can be derived by directly applying \autoref{eq:continuity} to yield the set of scalar non-equilibrium partial differential equations
\begin{subequations}
\begin{alignat}{3}
	\partial_{\tbar} [\rho_e]_1 &= 
		\Gammar\, \partial_{\rbar}^2 [\rho_e]_1 + \\
		&-c^2 \left( 1 - ([\rho_e]_2)^2\right)\arctanh{([\rho_e]_1)} + c\,\partial_{\rbar} [\rho_e]_2 \\
	\partial_{\tbar} [\rho_e]_2 &= 
		1\, \partial_{\rbar}^2 [\rho_e]_2 + \\
		&-\partial_{\rbar}\left(c \left( 1 - ([\rho_e]_2)^2\right)\arctanh{([\rho_e]_1)}\right).
\end{alignat}
\end{subequations}
These dynamic equations can be solved numerically. 
Moreover, it can be shown that the analytic steady state solution is
\begin{subequations}
\begin{align}
	[\rho_e]_1(\rbar) &= -\tanh{[\Omega_e]_1(\rbar_0)}\\
	[\rho_e]_2(\rbar) &= 
		-\tanh\parentheses*{[\Omega_e]_2(\rbar_0) + \frac{[\rho_e]_{2,\text{m}}}{[\rho_e]_{1,\text{m}}} \left( \phi(\rbar) - \phi(\rbar_0) \right)},
\end{align}
\end{subequations}
where $\rbar_0$ is the boundary coordinate used to determine boundary condition constants $[\Omega_e]_1(\rbar_0)$ and $[\Omega_e]_2(\rbar_0)$.


\section{Derivation of Dirac's Value Function from the Transport Model}\label{sec:DVF}

With the transport current \eqref{eq:j2} in hand, we are ready to derive Dirac's value function for systems described by the entropy of mixing.

\subsection{Separative Work and Separative Power}\label{sec:swu}

\begin{figure}
	\centering
	\includegraphics[width=1\linewidth]{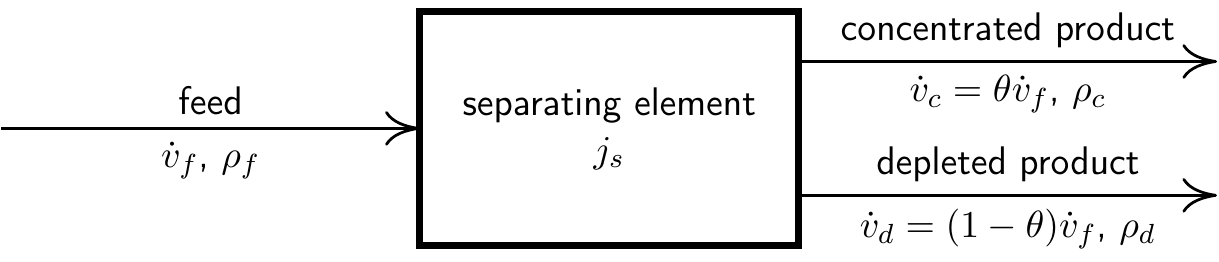}
	\caption{The diagram shows a feed flow $\vfd$ with quantity density $\rho_f$ entering a separating element with separation current $j_s$ that yields a concentrated product flow $\vcd$ with quantity density $\rho_c$ and a depleted product flow $\vdd$ with quantity density $\rho_d$.}
	\label{fig:separationTopology}
\end{figure}

Consider a separating element, shown in \autoref{fig:separationTopology}, that has a feed amount\footnote{The term ``amount'' here is intentionally ambiguous in order to allow the analysis to be as general as possible. The amount may be a mass, volume, or magnetic moment, for instance.} $v_f$ with quantity density $\rho_f$ entering.\footnote{The terminology here is that of separation science.\cite{Giddings1991,Lotkhov2005,Halvorsen2000}} The separating element outputs a concentrated product (with amount $v_c$ and quantity density $\rho_c$), and a depleted product (with amount $v_d$ and quantity density $\rho_d$). The flow rate of each of these amounts is denoted $\vfd$, $\vcd$, or $\vdd$.

Let $j_s$ be the separation current that flows between the concentrated and depleted products. It is $j_s$ that drives the separation. When $j_s = 0$, $\rho_f = \rho_c = \rho_d$, and no separation occurs.

Conservation requires that the amount entering the separating element must leave it; i.e.,
\begin{align}\label{eq:conservation}
	\vfd = \vcd + \vdd.
\end{align}
The product flow rates need not be equal. Let the fraction of the feed that becomes concentrated product be denoted $\theta$. This is commonly called the ``cut fraction.'' This, combined with \autoref{eq:conservation}, implies the following two equations:
\begin{align}
	\vcd &= \theta \vfd &\text{and} && \vdd &= (1-\theta) \vfd.
\intertext{
Another conservation that occurs is that of the quantity currents:
}
	\rho_c \vcd &= \rho_f \vcd + j_s &\text{and} && \rho_d \vdd &= \rho_f \vdd - j_s
\intertext{
The preceding four equations have eight variables, so four variables can now be found in terms of the other four. The following substitutions are useful as the derivation proceeds:
}
	\vcd &\mapsto \theta \vfd &\text{and} && 
	\rho_c &\mapsto \rho_f + \frac{j_s}{\theta \vfd}, \label{eq:subs1}\\
	\vdd &\mapsto (1-\theta) \vfd &\text{and} && 
	\rho_d &\mapsto \rho_f - \frac{j_s}{(1-\theta) \vfd}. \label{eq:subs2}
\end{align}

Let $F$ be a value function,\cite{Cohen1951} the form of which we will derive. The separative work $U$ is a measure of the amount of separation an element can provide, and, by assigning a value to the feed and products, is defined in terms of $F$ as\footnote{\label{fn:concentration}Most sources use concentration $C$ instead of quantity density $\rho$. The bijection that relates them is found from the equality $\rho = 2C - 1$.}
\begin{align}\label{eq:separativework}
	U = v_c F(\rho_c) + v_d F(\rho_d) - v_f F(\rho_f).
\end{align}
The separative power $\delta U$ is the rate at which separative work occurs, and can be written as
\begin{align}
	\delta U = \vcd F(\rho_c) + \vdd F(\rho_d) - \vfd F(\rho_f).
\end{align}
Applying the substitutions from Equations~\ref{eq:subs1} and \ref{eq:subs2}, this becomes
\begin{equation}
	\begin{split}
	\delta U &= \vfd \left( 
		\theta\, F\left(\rho_f + \frac{j_s}{\theta\vfd}\right) + \right. \\ 
		&\left.(1-\theta)\, F\left(\rho_f - \frac{j_s}{(1-\theta)\vfd}\right) - 
		F(\rho_f)
	\right).
	\end{split}
\end{equation}
For small separation currents $j_s$, the Taylor series expansion in $j_s$ about its origin yields zero for the zeroth and first orders. However, the second-order term gives the approximation
\begin{align}\label{eq:dU1}
	\delta U = \frac{j_s^2}{2 \vfd \theta (1-\theta)} \cdot \frac{d^2F(\rho_f)}{d\rho_f^2}.
\end{align}
This expression will now be used to derive Dirac's value function directly from the transport current $j_s$.

\subsection{Dirac's Value Function}\label{sec:dvf}

Let us restrict $F$ to functions that cause $\delta U$ to have no dependence on the feed quantity density $\rho_f$. From \autoref{eq:dU1}, then,
\begin{align}\label{eq:Fj}
	\frac{d^2F(\rho_f)}{d\rho_f^2} = \frac{1}{j_s^2}.
\end{align}
Given $j_s$, then, $F$ is specified to within two integration constants. If the separating element is governed by the transport model derived in Section~\ref{sec:frameworkandmodel}, its quantity density $[\rho_e]_2$ has current, from \eqref{eq:j2},
\begin{align}\label{eq:j3}
	j_s = c \left( 1 - ([\rho_e]_2)^2\right) \arctanh{([\rho_e]_1)} - \partial_{\rbar} [\rho_e]_2.
\end{align}
The first term in \eqref{eq:j3} describes separative flow and the second describes diffusive flow. However, if the gradient of $[\rho_e]_2$ is small\,---\,so that there is little separation occurring at each separative element\,---\,the separation current becomes
\begin{align}\label{eq:j4}
	j_s = c \left( 1 - ([\rho_e]_2)^2\right) \arctanh{([\rho_e]_1)}.
\end{align}
Inserting \eqref{eq:j4} into \eqref{eq:Fj} and integrating twice, the value function is
\begin{align}
	F([\rho_e]_2) = \frac{[\rho_e]_2}{2\, c^2 \arctanh^2 [\rho_e]_1} \arctanh [\rho_e]_2 + C_1 [\rho_e]_2 + C_0,
\end{align}
where $C_0$ and $C_1$ are constants of integration.
If we further require that $F(0) = 0$ (by convention) and $\left.dF/d[\rho_e]_2\right|_0 = 0$ (because, by symmetry, the value of $[\rho_e]_2$ should equal that of $-[\rho_e]_2$), the value function becomes
\begin{align}\label{eq:dvf1}
	F([\rho_e]_2) = \frac{[\rho_e]_2}{2\, c^2 \arctanh^2 [\rho_e]_1} \arctanh [\rho_e]_2,
\end{align}
which is Dirac's value function. Conventionally, the factor $2\, c^2 \arctanh^2 [\rho_e]_1$ is one-half, so we re-write \autoref{eq:dvf1} as 
\begin{align} \label{eq:dvf2}
	F([\rho_e]_2) = 2\, [\rho_e]_2 \arctanh [\rho_e]_2,
\end{align}
which is plotted in \autoref{fig:dvf}. See Appendix~\ref{app:dvf} for a discussion of the different forms of the DVF.

\begin{figure}
	\centering
	\includegraphics[width=1\linewidth]{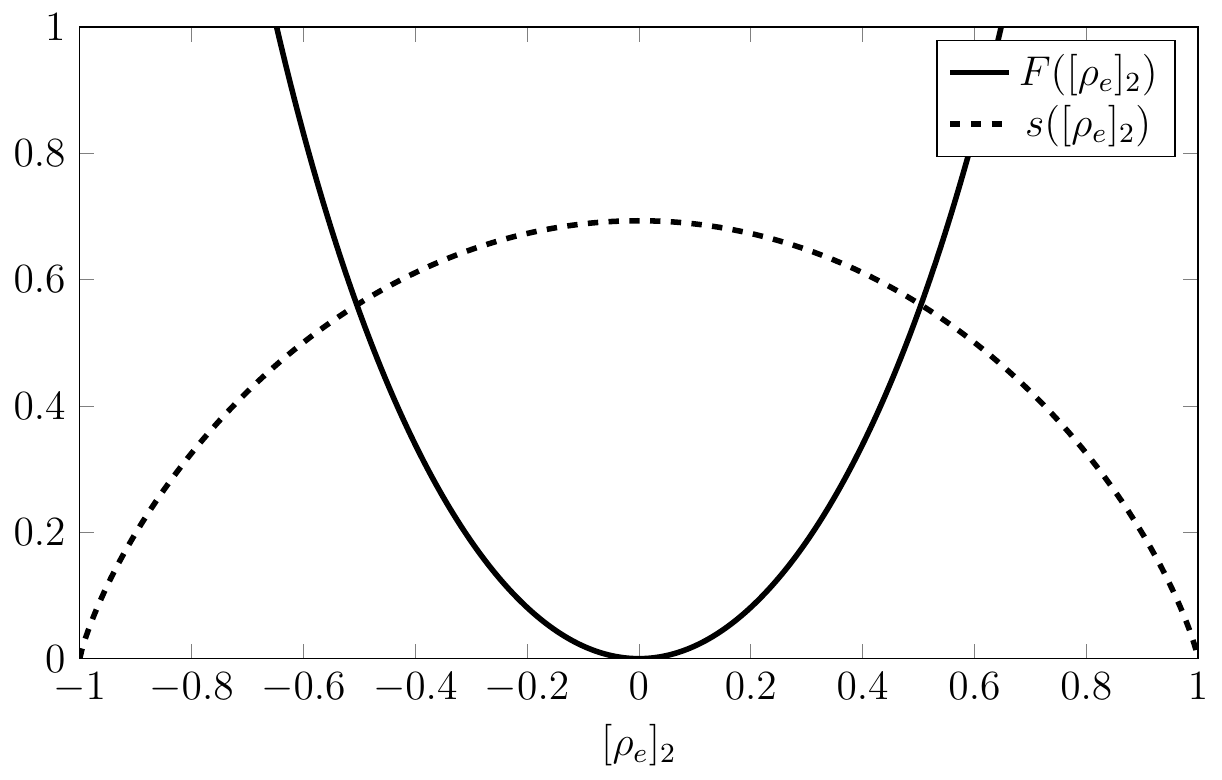}
	\caption{Dirac's value function $F$ \eqref{eq:dvf2} and the single-quantity entropy of mixing $s$. The value function is unbounded as its absolute value approaches unity.}
	\label{fig:dvf}
\end{figure}

\section{On the Validation of the Transport Framework and\\ the Generality of Dirac's Value Function}\label{sec:generality}

Now we have in-hand the elements needed to appreciate both the new validation of the non-equilibrium thermodynamical framework for transport analysis presented in Section~\ref{sec:frameworkandmodel} and the generality of Dirac's value function derived in Section~\ref{sec:DVF}. Deriving known results from a new theory is generally regarded as (partial) validation of the theory. It had been previously shown\cite{Picone2014b,Picone2015} that the framework for transport presented in Section~\ref{sec:framework} predicted several known results. However, these results did not include a connection to one of the key functions of isotope separation, the DVF. Such a connection would be expected of a transport framework of such supposed generality, and this connection has now been demonstrated in Section~\ref{sec:DVF}.

However, this validated generality when considered with the form of the derivation of Dirac's value function implies the generality of Dirac's value function, itself, which had not been previously appreciated in the literature. That is, the generality of its derivation implies the generality of its application. Previous derivations had limited the DVF to isotope and other mass transport applications. The derivation presented here has no such limitations, and yields a DVF that is applicable to any system described by the entropy of mixing.

To review, the following steps led us to the unique definition of Dirac's value function of Equation~\ref{eq:dvf2}:
\begin{enumerate}
	\item Derive a framework for the transport of any number of conserved quantity densities in any spatial configuration. (Section~\ref{sec:framework})
	\item Within the framework, define a one-dimensional, two-quantity model of transport for the transport quantity density $[\rho_e]_2$, with $[\rho_e]_1$ the local dynamical energy density. (Section~\ref{sec:model}) \label{it:model}
	\item Specify the entropy function to be the entropy of mixing $s$. (Section~\ref{sec:entropy})
	\item Derive the separative power function in terms of a separation current $j_s$. (Section~\ref{sec:swu})
	\item Require that the separative power function be independent of $[\rho_e]_2$, which, within two integration constants, specifies the value function $F$. (Section~\ref{sec:dvf})
	\item Require that $F(0) = 0$, by convention, in order to specify one integration constant. (Section~\ref{sec:dvf})
	\item Require that $\left.dF/d[\rho_e]_2\right|_0 = 0$, by symmetry, to specify the final integration constant. (Section~\ref{sec:dvf})
\end{enumerate}
This sequence is applicable to any system with separative transport that can be described by the transport model of Section~\ref{sec:model}. This is a broad class of systems that includes electrical, magnetic, inertial, isotopic, and ionic transport. To our knowledge, with the exception of isotope separation, Dirac's value function has not been applied in any of these fields. 
%

\section{Application to Spin Magnetization Transport}\label{sec:application}

We will now apply this finding to spin magnetization transport, which is of interest to the magnetic resonance imaging community.\cite{Picone2015,Picone2014b}

The transport equations for one-dimensional spin magnetization transport in a solid-state medium with a spatially varying magnetic field have been derived in the literature\cite{Picone2015} from the framework of Section~\ref{sec:framework}. This provides an example of a system to which the universalized Dirac value function can be applied.

This system is composed of spins in a solid medium, so the nuclei or electrons themselves do not transport, but the spin magnetization does. If all spins are aligned, the medium is completely magnetized or ``100 percent polarized.'' Conversely, if the spins are an equal mixture of ``up'' and ``down'' spins, the magnetization and polarization are zero.

In the presence of a background magnetic field $B$, polarization occurs in proportion to the field. Magnetic resonance technologies, which depend on this polarization, place their spin media in the strongest magnetic fields that can currently be realized. However, these fields produce polarization orders of magnitude lower than unity. A technique of hyperpolarizing media\,---\,that is, polarizing it beyond its thermal equilibrium\,---\,has been explored through the mechanism of dynamic nuclear polarization (DNP).\cite{Abragam1978,Griffin2010,Zotev2010,Krummenacker2012}

An alternative approach to hyperpolarizing a sample is spin magnetization transport, in which ``up'' spins are separated from ``down'' spins.\cite{Picone2015} In solids in high magnetic fields, spin magnetization is conserved by dipole-dipole exchange interactions. This perspective of magnetization transport allows the framework for transport of Section~\ref{sec:framework} to be applied.

\subsection{Magnetization Transport Model}

With reference to item~\ref{it:model} of the previous list, four physical insights are required to define a specific transport model:
\begin{enumerate}
	\item the identification of conserved quantities with densities $\rhovec$,
	\item the identification of the spatial manifold $\U$ with metric $g$,
	\item the selection of an entropy function $s$, and
	\item the transport rate tensor field $\Gammat$.
\end{enumerate}
We will consider one-dimensional transport with a Euclidean spatial metric $g = dr \otimes dr$ with $r$ a Cartesian coordinate.

With a spin system, there are two conserved quantities in $\rhovec$: (1) the total energy density, which is the sum of the Zeeman energy density (due to the spin-background magnetic field interaction) and the dipole energy density (due to the spin-spin magnetic field interaction) and (2) the spin magnetization.

As in Section~\ref{sec:bases}, it is useful to define an inhomogeneous basis. Let $B$ be a background magnetic field, let $B_d$ be the average spin-spin magnetic field, let $\mu$ be the magnetic moment of a single spin, and let $\Delta$ be the uniform spin density. The transformation to the inhomogeneous basis is given by the matrix
\begin{align}\label{eq:P2}
	P = 
	\begin{bmatrix}
		\dfrac{1}{B_d \mu \Delta} &
		-\dfrac{B}{B_d \mu \Delta} \\[2ex]
		0 &
		\dfrac{1}{\mu \Delta}
	\end{bmatrix}.
\end{align}

Under this transformation, $[\rho_e]_1$ represents the normalized dipole energy density\,---\,because the Zeeman energy density has been subtracted from the total\,---\,and $[\rho_e]_2$ represents the polarization (normalized magnetization).

The entropy density function $s$ of Section~\ref{sec:entropy} can be applied without modification. The transport rate tensor $\Gammat$ is system-dependent. For most spin systems, the two rates can be considered approximately equal,\cite{Genack1975} and therefore we let $[\Gamma_e] = [\Gamma_e]_1^1 = [\Gamma_e]_2^2$. An approximation of the transport rate is given by dimensional analysis to be
\begin{align}
	[\Gamma_e] = \frac{\mu}{4\pi} \hbar \gamma^2 \Delta^{1/3},
\end{align}
where $\gamma$ is a spin's gyromagnetic ratio and $\hbar$ is the Boltzmann constant.

A dimensionless spatial coordinate, temporal variable, and transport rate are used, as in Section~\ref{sec:dimensionless}. Applying the results of Section~\ref{sec:j}, we obtain the transport current
\begin{alignat}{2}\label{eq:j5}
	\jvec = &- \partial_{\rbar} [\rho_e]_1 \, e^1 \otimes d\rbar \nonumber \\
					&+ \left( c \left( 1 - ([\rho_e]_2)^2\right) \arctanh{([\rho_e]_1)} - \partial_{\rbar} [\rho_e]_2 \right) \, e^2 \otimes d\rbar,
\end{alignat}
where we have
\begin{align}
	c(\rbar) = \frac{1}{B_d} \cdot \frac{dB}{d\rbar}.
\end{align}
This transport current governs the flow of magnetization, both its diffusive and separative aspects. As we have shown in Section~\ref{sec:dvf}, this current yields a derivation of Dirac's value function.

\subsection{Dirac's Value Function and Separative Work for Magnetization}

It is now possible to regard the separative work $U$ of \autoref{eq:separativework}\,---\,with Dirac's value function $F$, given in \autoref{eq:dvf2}\,---\,as a measure of magnetization separation capability. This is a new valuation measure for emerging separative magnetization transport technologies.\cite{Picone2015,Picone2014b} Potential applications of these technologies are found in magnetic resonance imaging technologies from clinical scales (magnetic resonance tomography) to microscopic scales (magnetic resonance force microscopy). These technologies seek to hyperpolarize imaging samples through separative magnetization transport; the universalized version of Dirac's value function, presented here, provides a metric for designing and comparing magnetization separation elements. Moreover, by showing that Dirac's value function can be derived from the framework of transport analysis presented in Section~\ref{sec:framework}, this study has further validated the framework itself, and models derived therefrom.

\section*{Acknowledgement}
  This work was supported by the Army Research Office (ARO) MURI program under Contract \#~W911NF-05-1-0403. We also acknowledge the editorial contributions of Kayleen Rose Kondrack.

\bibliographystyle{abbrv}
\bibliography{./article}

\begin{thebibliography}{10}

\bibitem{Abragam1978}
A.~Abragam and M.~Goldman.
\newblock Principles of dynamic nuclear polarisation.
\newblock {\em Reports on Progress in Physics}, 41(3):395, 1978.

\bibitem{Bernstein2009}
J.~Bernstein.
\newblock Swu for u and me.
\newblock {\em ArXiv e-prints}, arXiv:0906.2505 [physics.hist-ph], 2009.

\bibitem{Casimir1945}
H.~B.~G. Casimir.
\newblock On onsager's principle of microscopic reversibility.
\newblock {\em Rev. Mod. Phys.}, 17:343--350, Apr 1945.

\bibitem{Chandler1987}
D.~Chandler.
\newblock {\em Introduction to modern statistical mechanics}.
\newblock Oxford University Press, New York, 1987.

\bibitem{Cohen1951}
K.~Cohen.
\newblock {\em The Theory of Isotope Separation as Applied to the Large-Scale
  Production of $\text{U}^{235}$}.
\newblock McGraw-Hill Book Company, first edition, 1951.

\bibitem{Denbigh1981}
K.~Denbigh.
\newblock {\em The Principles of Chemical Equilibrium: With Applications in
  Chemistry and Chemical Engineering}.
\newblock Cambridge University Press, 1981.
\newblock Section 8.1.

\bibitem{Dirac:1995aa}
P.~A.~M. Dirac.
\newblock The theory of the separation of isotopes by statistical methods.
\newblock In {\em The Collected Works of P.\ A.\ M.\ Dirac: 1924--1948}, pages
  1003--11. Cambridge University Press, 1995.
\newblock Unpublished notes, written \emph{circa} 1941.

\bibitem{Feynman2011}
R.~Feynman, R.~Leighton, and M.~Sands.
\newblock {\em The Feynman Lectures on Physics: Mainly electromagnetism and
  matter. Volume 2}.
\newblock Basic Books. Basic Books, 2011.

\bibitem{Genack1975}
A.~Z. Genack and A.~G. Redfield.
\newblock Theory of nuclear spin diffusion in a spatially varying magnetic
  field.
\newblock {\em Phys. Rev. B}, 12:78--87, Jul 1975.

\bibitem{Giddings1991}
J.~C. Giddings.
\newblock {\em Unified separation science}.
\newblock Wiley, New York, 1991.

\bibitem{Griffin2010}
R.~G. Griffin and T.~F. Prisner.
\newblock High field dynamic nuclear polarization-the renaissance.
\newblock {\em Phys. Chem. Chem. Phys.}, 12:5737--5740, 2010.

\bibitem{Halvorsen2000}
I.~J. Halvorsen and S.~Skogestad.
\newblock {\em Distillation Theory}, chapter~2.
\newblock Unknown, August 2000.

\bibitem{Hollands:2015aa}
S.~Hollands and R.~M. Wald.
\newblock Quantum fields in curved spacetime.
\newblock {\em Physics Reports}, 574:1--35, 2015.

\bibitem{Isaac2016}
C.~E. Isaac, C.~M. Gleave, P.~T. Nasr, H.~L. Nguyen, E.~A. Curley, J.~L. Yoder,
  E.~W. Moore, L.~Chen, and J.~A. Marohn.
\newblock Dynamic nuclear polarization in a magnetic resonance force microscope
  experiment.
\newblock {\em arXiv:1601.07253 [cond-mat.mes-hall]}, 2016.

\bibitem{Krummenacker2012}
J.~G. Krummenacker, V.~P. Denysenkov, M.~Terekhov, L.~M. Schreiber, and T.~F.
  Prisner.
\newblock Dnp in mri: An in-bore approach at 1.5 t.
\newblock {\em Journal of Magnetic Resonance}, 215(0):94 -- 99, 2012.

\bibitem{Landau2013}
L.~Landau and E.~Lifshitz.
\newblock {\em Fluid Mechanics}.
\newblock Elsevier Science, 2013.

\bibitem{Lebon2010}
G.~Lebon, D.~Jou, and J.~Casas-V\'{a}zquez.
\newblock {\em Understanding Non-equilibrium Thermodynamics: Foundations,
  Applications, Frontiers}.
\newblock SpringerLink: Springer e-Books. Springer Berlin Heidelberg, 2010.

\bibitem{Lee:2012ab}
J.~M. Lee.
\newblock {\em Introduction to Smooth Manifolds}.
\newblock Springer, second edition, 2012.
\newblock Problem 16-22.

\bibitem{Lee:2012aa}
J.~M. Lee.
\newblock {\em Introduction to Smooth Manifolds}.
\newblock Springer, second edition, 2012.

\bibitem{Lotkhov2005}
V.~Lotkhov, V.~Dil'man, A.~Lipatova, S.~Kvashnin, and N.~Kulov.
\newblock Profiles of the concentrations of components along the column height
  at different liquid holdup distributions in distillation of binary and
  ternary mixtures.
\newblock {\em Theoretical Foundations of Chemical Engineering}, 39(1):1 -- 4,
  2005.

\bibitem{Muller2014b}
I.~M\"{u}ller.
\newblock Entropy in nonequilibrium (chapter 5).
\newblock In A.~Greven, G.~Keller, and G.~Warnecke, editors, {\em Entropy},
  Princeton Series in Applied Mathematics, pages 79--104. Princeton University
  Press, 2014.

\bibitem{Ni2013}
Q.~Z. Ni, E.~Daviso, T.~V. Can, E.~Markhasin, S.~K. Jawla, T.~M. Swager, R.~J.
  Temkin, J.~Herzfeld, and R.~G. Griffin.
\newblock High frequency dynamic nuclear polarization.
\newblock {\em Accounts of Chemical Research}, 46(9):1933--1941, 2013.

\bibitem{Note1}
It is perhaps physically natural to consider the quantities $\protect \bm {\rho
  }$ to be dual to the potentials $\protect \bm {\Omega }$, as we have done
  here, although the choice is a matter of convention and the converse is also
  valid.

\bibitem{Note2}
The exterior derivative operator $d$ is here taken with respect to the vector
  space $V^*$.

\bibitem{Note3}
Some quantities can be negative, whereas others will use only the interval
  $[0,1]$.

\bibitem{Note4}
The term ``amount'' here is intentionally ambiguous in order to allow the
  analysis to be as general as possible. The amount may be a mass, volume, or
  magnetic moment, for instance.

\bibitem{Note5}
The terminology here is that of separation science.\cite
  {Giddings1991,Lotkhov2005,Halvorsen2000}.

\bibitem{Note6}
\label {fn:concentration}Most sources use concentration $C$ instead of quantity
  density $\rho $. The bijection that relates them is found from the equality
  $\rho = 2C - 1$.

\bibitem{Onsager1931}
L.~Onsager.
\newblock Reciprocal relations in irreversible processes. i.
\newblock {\em Phys. Rev.}, 37:405--426, Feb 1931.

\bibitem{Onsager1953}
L.~Onsager and S.~Machlup.
\newblock Fluctuations and irreversible processes.
\newblock {\em Physical Review}, 91:1505 -- 1512, 1953.

\bibitem{Picone2014b}
R.~A. Picone.
\newblock {\em Separative magnetization transport: theory, model, and
  experiment}.
\newblock PhD thesis, University of Washington, 2014.

\bibitem{Picone2015}
R.~A. Picone, J.~L. Garbini, and J.~A. Sidles.
\newblock Modeling spin magnetization transport in a spatially varying magnetic
  field.
\newblock {\em Journal of Magnetism and Magnetic Materials}, 374(0):440 -- 450,
  2015.

\bibitem{Prigogine1968}
I.~Prigogine.
\newblock {\em Introduction to thermodynamics of irreversible processes}.
\newblock Interscience Publishers, 1968.
\newblock p.\ 16.

\bibitem{Ruppeiner1979}
G.~Ruppeiner.
\newblock Thermodynamics: A riemannian geometric model.
\newblock {\em Phys. Rev. A}, 20:1608--1613, Oct 1979.

\bibitem{Ruppeiner1995}
G.~Ruppeiner.
\newblock Riemannian geometry in thermodynamic fluctuation theory.
\newblock {\em Rev. Mod. Phys.}, 67:605--659, Jul 1995.

\bibitem{Ruppeiner:2010lr}
G.~Ruppeiner.
\newblock Thermodynamic curvature measures interactions.
\newblock {\em American Journal of Physics}, 78(11):1170--80, 2010.

\bibitem{Sidles:2009cl}
J.~A. Sidles, J.~L. Garbini, L.~E. Harrell, A.~Hero, J.~P. Jacky, J.~R.
  Malcomb, A.~G. Norman, and A.~M. Williamson.
\newblock Practical recipes for the model order reduction, dynamical
  simulation, and compressive sampling of large-scale open quantum systems.
\newblock {\em New Journal of Physics}, 11(6):065002 (96pp), 2009.

\bibitem{Whitaker1986}
S.~Whitaker.
\newblock Flow in porous media i: A theoretical derivation of darcy's law.
\newblock {\em Transport in Porous Media}, 1(1):3--25, 1986.

\bibitem{Yablonsky2011}
G.~S. Yablonsky, A.~N. Gorban, D.~Constales, V.~V. Galvita, and G.~B. Marin.
\newblock Reciprocal relations between kinetic curves.
\newblock {\em EPL (Europhysics Letters)}, 93(2):20004, 2011.

\bibitem{Zia2009}
R.~K.~P. Zia, E.~F. Redish, and S.~R. McKay.
\newblock Making sense of the {L}egendre transform.
\newblock {\em American Journal of Physics}, 77:614--622, 2009.

\bibitem{Zotev2010}
V.~S. Zotev, T.~Owens, A.~N. Matlashov, I.~M. Savukov, J.~J. Gomez, and M.~A.
  Espy.
\newblock Microtesla \{MRI\} with dynamic nuclear polarization.
\newblock {\em Journal of Magnetic Resonance}, 207(1):78 -- 88, 2010.

\end{thebibliography}
\makeatother

\appendix

\section{Forms of Dirac's Value Function}\label{app:dvf}

Section~\ref{sec:dvf} presents two forms of the DVF, and here we explore their relationship to two others, including the standard form from the literature
\begin{align}
	F(c) = (2 C - 1) \log{\frac{C}{1-C}}
\end{align}
where $C \in [0,1]$ is the concentration. Applying the substitution rule $C \mapsto ([\rho_e]_2 + 1)/2$ and the identity $2 \arctanh{x} = \log(1+x) - \log(1-x)$, we obtain \eqref{eq:dvf2}.

Another interesting form of \eqref{eq:dvf1} in terms of the thermodynamic potential $\Omegavec$ \eqref{eq:Omegarho}, with the conventional scaling, is
\begin{align}
	F([\Omega_e]_2) = 2\, [\Omega_e]_2 \tanh{[\Omega_e]_2}.
\end{align}
We do not believe this form has previously been considered, yet it is expressed as a function of thermodynamic potential, which is fundamental to transport dynamics.

\section{Roles of the Entropy Function}
\label{section: entropy roles}

The readings of the preceding section inform our appreciation that the entropy function serves multiple roles in transport theory.

\subsection{Dynamical (Musical) Roles of the Thermometric Tensor}
\label{section: musical}

	The metric tensor of general relativity is associated to a fundamental dynamical law: particles (including photons) move on geodesic curves. In contrast, no such fundamental dynamical principle is known to be associated to the thermometric tensor $\boldsymbol{\mathcal{G}}$ (according to M\"{u}ller's survey\cite{Muller2014b} at least).  
	Still, in cases in which no universal dynamical law is known, $\boldsymbol{\mathcal{G}}$ can at least provide a starting postulate for such a law, by naturally specifying a proportionality of gradients in thermodynamic potentials to currents of conserved quantities (as in Eqs. \ref{eq:j}--\ref{eq:G} of this article).  Such proportional relations are known to mathematicians as \emph{musical isomorphisms}, by reason of the natural action of metric tensors in lowering (``$\flat$'') and raising (``$\sharp$'') the indices of tensor coefficients (for details, see \emph{e.g.} Lee\cite[p.~342]{Lee:2012aa}). 
	 
	``Musical'' transport models are not guaranteed to be rigorously correct, but they have the considerable virtue of not being obviously wrong, in the sense that musical transport relations scrupulously respect the Four Laws.  
	By virtue of their (relative) ease of derivation, such relations can be a useful first step toward verification by experiment and/or optimization with the help of microscopic computational simulation. 

\subsection{Roles of the Riemann, Ricci, and Sectional Curvatures}
\label{section: curvature}

	We have not, in the present article, grappled with the many intricate and difficult issues that are associated to transport phenomena near phase transitions (here ``near'' can mean either or both of spatial distance or in thermodynamic potential-distance).  
	Ruppeiner\cite{Ruppeiner:2010lr} has argued (convincingly we think) that singularities in thermometric curvature are associated to the divergences in spatial correlation length that are characteristic of phase transitions.

	In transport theory the spatial metric tensor $g$ and the thermometric tensor $\boldsymbol{\mathcal{G}}$ occur jointly, and it is natural to wonder about the intrinsic geometric properties of the combined metric (including for example the sectional, Ricci, and Riemann curvatures).  One consideration is that general relativity is covariant under arbitrary coordinate transformations, whereas the natural invariance of transport equations is associated to affine transforms of the conserved quantity-densities (that is, arbitrary linear transforms and origin-shifts).  
	
	There is no shortage here of opportunities for further investigations regarding the interplay of spatial metric curvature (determined by $g$) and thermometric curvature (determined by $\boldsymbol{\mathcal{G}}$).  We consider Robert Wald's remarks in regard to the merits of mathematical abstraction in general relativity to be applicable also in transport theory.

\subsection{Physical Intuitions in Regard to Thermometric Tensors and Entropy}
\label{section: yoga}

	Our guiding physical insight\,%
	in regard to thermometric tensors and entropy functions is that the thermometric tensor $\boldsymbol{\mathcal{G}}$ originates fundamentally in Boltzmann fluctuations (whether experimentally observed or computationally simulated), such that the entropy function is well-defined if and only if the thermometric tensor (regarded as the Hessian of the entropy) is \emph{integrable}.  
	Further associated to this insight is the notion that Boltzmann fluctuations are \emph{measured} quantities, and therefore are properly realized (both in experiments and in computational simulations) as \emph{Lindblad processes}.\cite{Sidles:2009cl}

	Here too there is no shortage of open problems and opportunities for further investigation and unification, to which the survey of Holland and Wald\cite{Hollands:2015aa} provides an introduction.

\end{document}